\documentstyle[aps,epsf]{revtex}

\input{epsf}
\begin{document}


\title{The problem of Coulomb interactions in the theory of the quantum Hall 
effect.}

\author{M.A. Baranov$^{1,2}$, A.M.M. Pruisken$^{2}$, B. Skoric$^{2}$}
\address{
{\em (1)}{\ RRC ''Kurchatov Institute'', Kurchatov Square 1, 123182 Moscow, 
Russia}\\
{\em (2)} {\ University of Amsterdam, Valckenierstraat 65, 1018XE Amsterdam, 
The Netherlands}}
\date{\today }
\maketitle

\begin{abstract}
We summarize the main ingredients of a unifying theory 
for abelian quantum Hall states. This theory combines the Finkelstein 
approach to 
localization and interaction effects with the topological concept of an 
instanton vacuum as well as Chern-Simons gauge theory.
We elaborate on the meaning of a new symmetry ($\cal F$ invariance) for 
systems with an infinitely
ranged interaction potential. We address the renormalization of the theory and
present the main results in terms of a scaling diagram of the conductances.
\end{abstract}

\section{Introduction.}

In these notes  we discuss some of the recent 
advancements in the theory of the 
quantum Hall effect\cite{mishI,mishII,mishIII}. In particular, we address 
some of the main steps in the development
of a theory\cite{PrB} that combines the {\em instanton vacuum} approach for 
spin polarized, free 
electrons\cite{fp} with the {\em Finkelstein} treatment of the Coulomb 
interactions\cite{Fink} in the disorderd systems. 

The electron gas with an infinitely ranged interaction potential is, in many ways, very 
different from what we know about the theory of free electrons. This class of problems 
belongs to a different universality class of quantum transport phenomena and it is 
characterized by a typical interaction symmetry which we name $\cal F$ invariance\cite{mishI}. 
$\cal F$ invariance is intimidly related to the electrodynamic $U(1)$ gauge invariance and 
it has major consequences for the renormalization of the theory\cite{mishII}. 

The main physical objective of our theory is to unify the different aspects of (abelian) 
quantum Hall states that have originated from  different sources and that have been studied 
over the years independently. They include the quantum critical behavior of the quantum Hall 
plateau transitions\cite{exp}, composite fermion theory or the Chern Simons mapping between 
integral and fractional quantum Hall states\cite{qhecf}, the Luttinger liquid theory of 
quantum Hall edge excitations\cite{fracedge}, 
as well as the stability or robustness of the quantization phenomenon due to the disorder
\cite{PrB2}. For a
detailed exposure we refer the reader to the literature and here, we only present a brief 
introduction to the subject.

\section{Matrices in frequency and replica space}

Diffusive modes are encoded in the unitary matrix field variables $Q_{nm}^{\beta \gamma }$
\cite{Fink}. 
Here, the superscripts represent the replica indices
($\beta $,$\gamma =1,2,\ldots N_{r}$ where $N_{r} \rightarrow 0$ at the end of all calculations)
and the subscripts denote the Matsubara frequency indices 
($n,m=0,\pm 1,\pm 2,\ldots \pm N_{{\rm \max }}^{\prime }$ where the cut-off 
$N_{{\rm \max }}^{\prime }$ 
is sent to infinity in the end).
The matrices $Q$ generally obey the constraints

\[
Q^{\dagger }=Q,\ Q^{2}=1,\ {\rm tr}Q=0 
\]
and they can be represented by $Q= T^{\dagger} \Lambda T$ 
where $T$ is unitary and $\Lambda$ is a diagonal matrix

\[
\left( \Lambda \right) _{nm}^{\beta \gamma }=\delta ^{\beta \gamma }\delta
_{nm}{\rm sign}(m). 
\]
In order to fascilitate a discussion of the electrodynamic $U(1)$ gauge invariance of 
the theory, we generally follow a very specific cut-off procedure in frequency space and a very
specific set of algebraic rules for matrix manipulation which we term ${\cal F}$ algebra\cite{mishI}.
For example, if we write the matrix $Q$ in the form $Q=\Lambda +\delta Q$ then the
$\delta Q$ matrix is generally taken as a {\em small} matrix in frequency space, 
i.e. its matrix elements are non zero only for $n,m=0,\pm 1,\pm 2,\ldots \pm N_{
{\rm \max }}$, where $N_{{\rm \max }}\ll N_{{\rm \max }}^{\prime }$. In different
words, the unitary rotation $T$ mixes amongst (positive and negative) frequencies with
a {\em small} cut-off ($N_{max}$) whereas the $U(1)$ gauge transformations generally
involve {\em large} matrices with a {\em large} cut-off ($N_{{\rm \max }}^{\prime }$)
in frequency space. 

By employing distinctly different cut-offs $N_{{\rm \max }}$ and $N_{{\rm \max }}^{\prime }$ in 
Matsubara frequency space, both of which are sent to infinity in the end, the problem of 
electrodynamic gauge invariance simplifies dramatically\cite{mishI}. Physically, the cut 
off procedure is motivated by the vastly different 
energy scales that generally characterize the elastic scatering processes ($1/\tau_{el}$) on 
the one hand, and the bandwidth of the electron gas on the other. However, the rules of 
$\cal F$ algebra can be shown 
to have a quite universal significance for disordered electron systems. For example, it  
successfully descibes the dynamics of chiral edge excitations in quantum Hall systems. It has been
used as a microscopic basis for deriving, from first principles, the complete Luttinger 
liquid theory of edge excitations for abelian quantum Hall states\cite{mishIII,fracedge}.

We generally need the introduction of two more ({\em large}) matrices . First, 
the diagonal matrix $\eta$

\[
\left( \eta \right) _{nm}^{\beta \gamma }=\delta ^{\beta \gamma }m\delta
_{nm} 
\]
which is the matrix representation of (imaginary) time derivative. Secondly, there are
the off diagonal matrices $I_n^\alpha$

\[
\left( {\rm I}_{n}^{\alpha }\right) _{kl}^{\beta \gamma }=\delta ^{\alpha
\beta }\delta ^{\alpha \gamma }\delta _{n,k-l}\  
\]
which are the generators of the $U(1)$ gauge transformations. For more details on the
rules of $\cal F$ algebra and various algebraic identities, we refer the reader to the 
original papers \cite{mishI,mishII,mishIII}.

\section{The $\sigma$ model action}

The action consists of three terms\cite{mishI}

\begin{equation}
S[Q]=S_{\sigma }[Q]+S_{{\rm F}}[Q]+S_{{\rm C}}[Q] .  \label{1}
\end{equation}
Here, the first term

\begin{equation}
S_{\sigma }[Q]=\frac{\sigma _{xx}^{0}}{8}{\rm Tr} \left(
\partial _{\mu }Q\right) ^{2} -\frac{\sigma _{xy}^{0}}{8}
\varepsilon _{\mu \nu }{\rm Tr} Q\partial _{\mu }Q\partial _{\nu
}Q  \label{2}
\end{equation}
represents the standard {\em non linear $\sigma $ model}\cite{fp} for spinless, free 
electrons in two dimensions and in
the presence of a static perpendicular magnetic field. The
{\rm Tr} symbol stands for both the spatial integration and the trace {\rm tr} over all
matrix indices (Matsubara as well as replica). 

The second term $S_F$ is the {\em singlet interaction} term first introduced by 
Finkelstein\cite{Fink}. It can be written in three equivalent ways\cite{mishI}

\begin{eqnarray}
S_{{\rm F}}[Q] &=&-\pi z_{0}T\int d^{2}x\left[ \sum_{\alpha
,n_{i}}Q_{n_{1}n_{2}}^{\alpha \alpha }Q_{n_{3}n_{4}}^{\alpha \alpha }\delta
_{n_{1}+n_{3},n_{2}+n_{4}}+4{\rm tr} \eta Q \right] +{\rm const}
\nonumber \\
&=&-\pi z_{0}T\int d^{2}x\left[ \sum_{\alpha ,n}{\rm tr} {\rm I}
_{n}^{\alpha }Q {\rm tr} {\rm I}_{-n}^{\alpha }Q +4{\rm 
tr}\eta Q\right] +{\rm const}  \nonumber \\
&=&-\frac{\pi }{2}z_{0}T{\rm Tr} \bigl[ {\rm I}_{n}^{\alpha
},Q\bigr] \bigl[ {\rm I}_{-n}^{\alpha },Q\bigr] .  \label{3}
\end{eqnarray}
Here, the quantity $z_0$ represents the singlet interaction amplitude and $T$ is the 
temperature. The compact
notation of the last line indicates that the expression is invariant under $U(1)$
gauge transformations ($\cal F$ invariance, see Section 4). This expression generally
acts as a single operator under renormalization group transformations\cite{PrB,mishII}.

The non linear $\sigma$ model part and $S_F$ represent, together, the effective action
for a system with infinitly ranged electron-electon interactions. 
The Coulomb potential appears explicitly only in higher dimensional (irrelevant) terms
($S_C$) which are usually discarded. However, these higher dimensional operators 
turn out to be {\em dangerously } irrelevant and it is generally important to also take 
the term $S_C$ (the socalled Coulomb term) into account. This part of the action can be 
written as

\begin{equation}
S_{{\rm C}}[Q]=\pi T\int d^{2}xd^{2}x^{\prime }\sum_{\alpha ,n}{\rm tr}
 {\rm I}_{n}^{\alpha }Q({\bf x}) U^{-1}({\bf x}-{\bf x}^{\prime
}){\rm tr} {\rm I}_{-n}^{\alpha }Q({\bf x}^{\prime }),
\label{4}
\end{equation}
where

\[
U^{-1}({\bf p})=\int d^{2}xU^{-1}({\bf x})e^{-i{\bf px}}=\frac{\pi }{2}\ 
\frac{1}{1+\rho U_{0}({\bf p})}\approx \Gamma \ |{\bf p}| 
\]
In this expression
$\rho =dn/d\mu $ represents the thermodynamic density of states and $U_{0}({\bf p}%
)=2\pi e^{2}/|{\bf p}|$ is the bare Coulomb interaction in two dimensions.

\subsection{Renormalization}
The theory, as it stands, contains only four quantities for which one needs to compute
the quantum corrections, i.e. $\sigma_{xx}^0$, $\sigma_{xy}^0$, $z_0$ and $\Gamma$.
As is well known, the quantity  $\sigma_{xy}^0$, multiplying the topological charge $q$

\begin{equation}
q(Q)=\frac{1}{16\pi i} Tr \epsilon_{\mu\nu} Q\partial_\mu Q \partial_\nu Q 
=\frac{1}{4\pi i} \oint dx tr T\partial_x T^{\dagger}
\Lambda ,
\end{equation}
is not affected by the perturbative quantum theory and will be dealt with at a later stage.
The quantity $\Gamma$, the other hand, remains strictly unrenormalized and this statement,
as it has turned out from the microscopic derivation of the action, should be imposed as a general 
constraint on the quantum theory\cite{PrB2} (see also Section 7).

This leaves us with two non trivial renormalizations in perturbation theory, i.e. the 
inverse coupling constant $\sigma_{xx}^0$ and the interaction amplitude $z_0$.
The complete list of renormalization group $\beta$ and $\gamma$ functions in $2+2\epsilon$
dimension is given by\cite{mishII}

\begin{eqnarray*}
\beta _{t} &=&\frac{dt}{d\ln L}=-2\varepsilon t+2t^{2} +O(t^3 ) , \\
\beta _{xy} &=&\frac{d\sigma _{xy}}{d\ln L}=0, \\
\gamma  &=&\frac{d\ln z}{d\ln L}=-t-t^{2}(3+\pi ^{2}/6) +O(t^3 ) , \\
\gamma _{C} &=&\frac{d\ln \Gamma }{d\ln L}=-1.
\end{eqnarray*}
Here, we have written $t=\frac{1}{\pi \sigma _{xx}}$ and $L$ denotes the linear dimension of 
the system. These results are quite similar to those of the classical Heisenberg ferromagnet
and the physics of the electron gas in $2+2\epsilon$ dimension can be obtained
following the (in many ways) unique field theoretic methodology of dealing with Goldstone modes. 

\subsection{Fermi liquids versus Non-Fermi liquids}
It is important to keep in mind, however, that the physics of interacting systems is 
very different from that of free electron localization. Free electron formalisms,
unlike the Finkelstein formalism, has $Q$ martrix field variables which usually have
two frequencies only, i.e. the advanced and retarded ones\cite{fp}. 
A formal but general
way of describing the crossover between the single particle and many body formalisms is 
obtained by varying the
frequency cut-off $N_{max}$ in the $Q$ matrix fields, from unity to infinity.
By varying the value of $N_{max}$, the theory changes fundamentally.
The most dramatic effect of putting $N_{max} \rightarrow \infty$ is that
the ultraviolet singularity structure of theory (i.e. the $\beta$ function) completely 
changes\cite{PrB}.
The presence of the singlet interaction term $S_F$ now implies that the problem generally
belongs to a different universality class of quantum transport phenomena. Since there is no 
Fermi liquid principle for the {\em disordered} electron system with an infinite range 
interaction potential,  
it is necessary to reconsider the topological concept of a $\theta$ or instanton vacuum which
previously was introduced and investigated for the free electron theory alone\cite{fp,inst}.

Along with the renormalization behavior, also the structure of the operators of the theory 
change as the cut off $N_{max}$ is being sent to infinity\cite{mishII}. 
A new, previously unrecognized 
notion of {\em interaction symmetries} now becomes an integral part of the problem. 
These symmetries ($\cal F$ invariance, Section 4) 
are intimidly related to the electrodynamic $U(1)$ gauge invariance and much is yet
to be learned about the behavior of the theory in the strong coupling regime.

Before elaborating on symmetries and gauge invariance, however, we wish to first address some 
of the general ideas that are associated with the perturbative renormalization group results 
of the previous Section.

Notice that on the basis of the $\beta$ function or asymptotic freedom alone, one generally 
expects that the interacting electron gas, in two spatial dimensions, behaves 
{\em quasi metallic} at short distances but it eventually enters a 
strong coupling phase with a massgap (an {\em insulating} phase), as the lengthscale is increased. 

The renormalization of the interaction terms $S_F$ and $S_C$, i.e. the $\gamma$ and 
$\gamma_c$ functions, generally determine the {\em dynamical} scaling in the problem,
i.e. the temperature and frequency dependence of physical obsevables, and this includes 
the non-Fermi liquid behavior of quantities like the specific heat\cite{mishII}.

The result of the $\gamma$ indicates that the singlet interaction term $S_F$ plays formally 
the role of an {\em order parameter} (i.e. the spontaneous magnetization) in the context of
conventional critical phenomena phenomenology. Physically it means that the theory generates 
a (Coulomb) gap in the density of states that enters in the expression for the specific 
heat\cite{mishII}.
This result is quite different from what one is
used to in the theory free electron localization, or in the theory with a finite value of $N_{max}$. 
For example, such free particle concepts like anomalous or {\em
multi fractal} fluctuations in the local density of states are no longer valid in the theory with 
Coulomb interactions. The physics of the $\gamma$ functions is generally very different.

\begin{figure}
  \centerline{
  \epsfbox{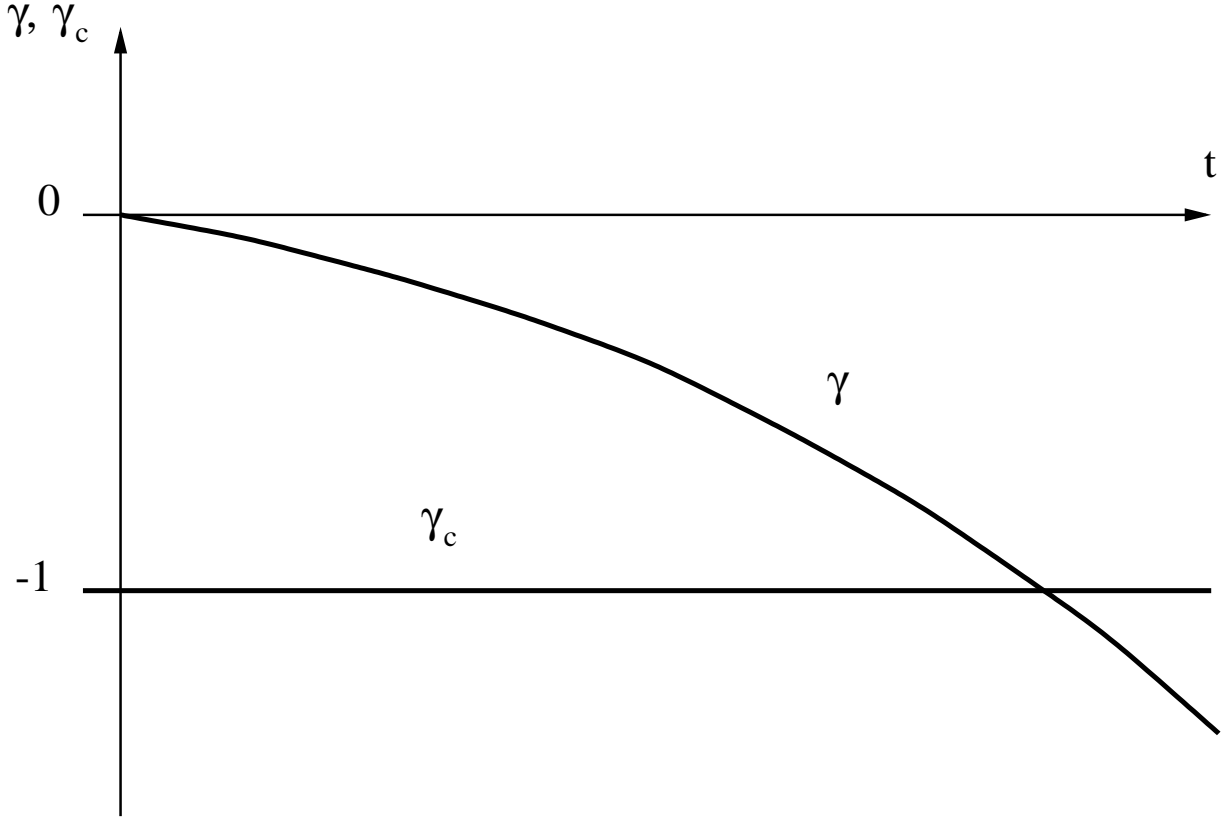}}
\caption{The anomalous dimensions $\gamma$ and $\gamma_c$ versus $t$.}
\end{figure}

In Fig. 1 we have plotted the results for the anomalous dimensions $\gamma$ and 
$\gamma_c$ versus $t$. 
We see that in the weak coupling or small $t$ regime, the $\gamma (t)$ dominates the $\gamma_c (t)$
indicating that the Coulomb term $S_C$ is irrelevant. However, with the $\gamma_c (t)$ 
function remaining fixed
at the value $-1$, as mentioned before, there is likely to be a point on the $t$ axis beyond which 
the $\gamma_c (t)$ dominates the $\gamma (t)$ function. This means that upon entering the strong 
coupling regime, the roles of $S_F$ and $S_C$ get interchanged and the dynamics of the 
insulating phase is now entirely determined by the Coulomb term $S_C$.

This dangerously irrelevant behavior of $S_C$ has not been recognized previously but obviously 
plays a fundamental role in the theory of metals and insulators. 
This notion can not be obtained in any heuristic
or phenomenological fashion, but it clearly affects the way in which we are going to look upon
the complications of the theory in dealing with the
quantum Hall effect. For example, if the quantum critical singularities
of the quantum Hall plateau transitions\cite{exp} are appropriately described by the 
non-perturbative  behavior of the $\beta_{xy}$ and $\gamma$ functions of the 
Finkelstein theory\cite{PrB}, then this
critical behavior is certainly very different from the main expectations of the free particle
approximation. Unlike the free particle problem which effectively lives in two spatial
dimensions alone, the Finkelstein theory contains the time variable as an extra non-trivial
dimension. This not only destroys any hope of finding an exact (conformal) scheme of 
critical indices, but also invalides any explicit or implicit attempt of employing Fermi liquid 
ideas for quantum Hall systems in the presence of the Coulomb interactions.

\section{{$\cal F$} invariance}

{$\cal F$} invariance \cite{mishI} is one of the most important interaction symmetries of 
disordered systems. It means that the action for systems with an infinite range interaction 
potential (like the Coulomb potential) is invariant under spatially independent $U(1)$ gauge
transformations. Such gauge transformations can be represented by a ({\em large}) unitary
matrix $W$

\[
W=\exp (i\sum_{\alpha ,n} \phi_{n}^{\alpha }{\rm I}_{n}^{\alpha }) 
\]
where the $\phi_{n}^{\alpha }$ are the frequency components of the imaginary time quantity 
$\phi (\tau)$. 
The statement of $\cal F$ invariance can now be written as

\[
S[Q]=S[W^{-1}QW]. 
\]
$\cal F$ invariance is generally violated by systems with a finite range interaction potential, 
or free electron systems. One can show that these systems map, under the action of the 
renormalization group, back onto the free electron theory and, therefore, Fermi liquid ideas 
can be applied in this case.

The concept of $\cal F$ invariance implies that only the quantities and
correlations that are $\cal F$ invariant have a simple infrared behavior that generally
can be handled with the methodology of the renormalization group. It also implies
that the {\em dynamical scaling} of physical observables like the conductances
can only be extracted from the theory if $\cal F$ invariance is respected by the
renormalization scheme that one chooses. For example, the momentum shell or
background field methodology generally violate $\cal F$ invariance and this 
complicates the computation of e.g. the AC conductances. 

On the other hand,
the partition function itself, or the response to electromagetic fields,
generally respects statement of $\cal F$ invariance. It is therefore 
important to study $\cal F$ invariant quantities in general and see $\cal F$ algebra at work.

\section{External EM fields}

If, for the sake of simplicity, we first consider the theory for weak static magnetic fields, 
then the various pieces of the action, in the presence of scalar and vector potentials,
can be written in a transparant fashion as follows\cite{mishII}

\begin{eqnarray*}
S_{\sigma }[Q,A]&=&\frac{\sigma
_{xx}^{(0)}}{8}{\rm Tr}\left[ D_{\mu },Q\right] ^{2} -\frac{
\sigma _{xy}^{(0)}}{8}\varepsilon _{\mu \nu }{\rm Tr} Q\left[ D_{\mu
},Q\right] \left[ D_{\nu },Q\right]  , \\
S_{{\rm F}}[Q,A] &=& S_{{\rm F}}[Q], \\ 
S_{{\rm C}}[Q,A]&=&\pi T\int \frac{d^{2}p}{(2\pi
)^{2}}\sum_{\alpha ,n}\left[ {\rm tr}({\rm I}_{n}^{\alpha }Q)-\frac{1}{\pi T}
\left( A_{0}\right) _{-n}^{\alpha }\right] U^{-1}({\bf p})\left[ {\rm tr}
({\rm I}_{-n}^{\alpha }Q)-\frac{1}{\pi T}\left( A_{0}\right) _{n}^{\alpha
}\right] . 
\end{eqnarray*}
Here, $D_{\mu }=\partial _{\mu }-iA_{\mu }$ is the covariant derivative in matrix form with 
$A_{\mu }=\sum_{\alpha ,n}(A_{\mu })_{n}^{\alpha }{\rm I}_{n}^{\alpha }$.
By the rules of $\cal F$ algebra, the $U(1)$ gauge invariance can now be formulated by saying that
the following set of transformations
$Q \rightarrow  W^{-1} Q W $, 
$( A_\mu )_n^\alpha \rightarrow ( A_\mu )_n^\alpha +i \partial_\mu \phi_n^\alpha $, and
$( A_0 )_n^\alpha \rightarrow ( A_0 )_n^\alpha +i \omega_n \phi_n^\alpha$ 
leaves the action invariant.

The theory for strong static magnetic fields $B$ is slightly more complex and has additional
terms ($\sigma_{xy}^{II}$, Section 7) reflecting the $B$ dependence of the electron density.

\section{Response at a tree level}

As a simple check of the above formulae, we compute the gauge invariant electromagnetic
response at a tree level, for the case $\sigma _{xy}=0$. We obtain

\begin{equation}
S_{{\rm eff}}^{tree} [A]= \frac{1}{T}\sum_{\alpha ,n}\int \frac{%
d^{2}p}{(2\pi )^{2}}\frac{\sigma _{xx}^{(0)}}{8}\left[ \delta _{\mu \nu }-%
\frac{p_{\mu }p_{\nu }}{p^{2}+4\omega _{n}U^{-1}(p)/\sigma _{xx}^{(0)}}%
\right] \frac{(E_{\mu })_{n}^{\alpha }\overline{(E_{\nu })_{n}^{\alpha }}}{%
\omega _{n}}.  \label{7}
\end{equation}
where $(E_{\mu })_{n}^{\alpha }$ is the electric field. 
The charge density can be defined as $n_{n}^{\alpha }(p) 
=-T\delta S_{{\rm eff}}[A]\delta \left(A_{0}\right) _{-n}^{\alpha }$,
then from $S_{{\rm eff}}^{tree} [A]$ one obtains the continuity equation that in ordinary space 
time notation can be written as\cite{mishII}

\begin{equation}
\partial _{t}n+{\bf \nabla} \cdot ({\bf j}_{{\rm C}}+{\bf j}_{{\rm diff}})=0,
\label{8}
\end{equation}
where
$
{\bf j}_{{\rm diff}}=-D_{xx}^{(0)}\nabla n\ 
$
with $D_{xx}^{(0)}=\sigma _{xx}^{(0)}/2\pi \rho $ 
is just diffusive current component and
$
{\bf j}_{{\rm C}}=\frac{\sigma _{xx}^{(0)}}{2\pi }{\bf E}_{{\rm tot}}\ 
$
with 
$
{\bf E}_{{\rm tot}}={\bf E}_{{\rm ext}}-\nabla \int d^{2}x^{\prime
}U_{0}({\bf x}-{\bf x}^{\prime })n({\bf x}^{\prime }) 
$
is the electric current density due to the external and
internally generated electric fields.
The tree level response therefore reproduces the
well known results of the theory of metals.

\section{Response with quantum corrections}

We have computed the complete response to extenal electromagnetic fields with
quantum corrections and checked the gauge invariance at a one loop level. 
The analysis is rather lengthy and the details will be reported elsewhere\cite{PrB2}.
We present, instead, the final result for the continuity equation which
can be written in frequency and momentum notation as follows

\begin{equation}
\omega _{\eta } ( n_{\eta } -  \frac{\sigma
_{xy}^{(0)}}{2\pi } B^\eta ) =ip_{\mu }\left\{ \frac{\sigma _{xx}^{{\rm l}}}{2\pi 
}\left( E_{\mu }^{\eta }-ip_{\mu }U_{0}n_{\eta }\right)- D_{xx} iq_{\mu
}\left( n_{\eta }+i\frac{\sigma _{xy}^{{\rm II}}}{2\pi }B^{\eta }\right)
\right\} . \label{9}
\end{equation}
Here, $\sigma _{xx}^{{\rm l}}$ is the {\em renormalized} longitudinal conductivity which is
expressed as $\sigma_{xx}^{(0)}$ plus quantum corrections. 
Furthermore, $D_{xx}=\sigma _{xx}^{{\rm l}}/2\pi \rho$ the renormalized diffusive coefficient 
and $\sigma_{xy}^{{\rm II}}/2\pi =dn/dB$. 

It is important to remark that the static ($\omega_\eta =0$) limit of this expression has 
an important general significance for the renormalization of the theory. 
Notice that by putting $\omega_{\eta}=0$, the expression only
contains the thermodynamic quantities such as $\rho$, the thermodynamic density
of states, $\sigma_{xy}^{\rm II} = \partial n / \partial B$, which is the derivative of 
the electron density with respect to the static external $B$, as well as the bare Coulomb 
interaction $U_0$. This form of the static response can be shown to be quite generally 
valid, independently of the effective action of the $Q$ field variables. This means that 
the thermodynamic quantities  $\rho$, $\sigma_{xy}^{\rm II}$ as well as $U_0$ generally 
do not acquire any quantum corrections and this statement can be imposed as an
important general constraint on the quantum theory.

Notice that this constraint does not involve the quantities $\sigma_{xx}$, $\sigma_{xy}$ and the 
interaction amplitude $z$ 
which do not appear in the expression for static response. These quantities are therefore the only 
ones for which
quantum corrections are possible (and do occur) in general.

\section{Physical observables}

The results of the previous Section imply that a general quantum theory of physical observables
can be formulated and expressed in terms of $\cal F$ invariant correlations of the $Q$ field 
variables.
For example, the linear response to an external electric field can be generally expressed 
in terms of a quantity $\sigma_{xx}^{\prime}$ (Kubo formula) as follows 

\begin{eqnarray*}
\sigma _{xx}^{\prime } &=&-\frac{\sigma _{xx}^{(0)}}{4\eta }\left\langle 
{\rm tr}\left\{ \left[ {\rm I}_{\eta }^{\alpha },Q\right] \left[ {\rm I}%
_{-\eta }^{\alpha },Q\right] \right\} \right\rangle +\frac{\sigma
_{xx}^{(0)2}}{16\eta D}\int_{x^{\prime }}\left\langle {\rm tr}\left\{ \left[ 
{\rm I}_{\eta }^{\alpha },Q\right] \partial _{\mu }Q\right\} {\rm tr}\left\{
\left[ {\rm I}_{-\eta }^{\alpha },Q\right] \partial _{\mu }Q\right\}
\right\rangle \\
\end{eqnarray*}
This result, when evaluated perturbatively \cite{mishII}, is of the general form 
$\sigma _{xx}^{(0)}+{\rm quantum\ corrections}$.
A similar expression exist for the Hall conductance $\sigma_{xy}^{\prime}$. These general results retain 
their significance also beyond the theory of perturbative quantum corrections. For example, they can be used for 
non-perturbative (instanton) calculus \cite{PrB} as well and this has led to the previously unrecognized 
concept of $\theta$ renormalization, or renormalization of the Hall conductance $\sigma_{xy}$ \cite{fp}.

For completeness, we mention that the list of effective parameters $\sigma_{xx}^{\prime}$ and $\sigma_{xy}^{\prime}$
can also be extended to include an effective quantity $z^{\prime}$ which is associated with the interaction
amplitude $z_0$. The result can be expressed as \cite{mishII}

\[
\sum_{n>0}\omega _{n} z^{\prime} (\omega_n ) =
\frac{\pi }{2}%
z_{0} T \sum_{n>0}\left\langle {\rm tr} 
\bigl[ 
{\rm I}_{n}^{\alpha
},Q
\bigr] \bigl[ 
{\rm I}_{-n}^{\alpha },Q
\bigr] 
\right\rangle 
\]

\section{Instantons}

The non-perturbative contributions from a topologically nontrivial sectors of 
the theory (instantons) have formally the same structure as 
those obtained in the theory of free electrons \cite{inst}. Since the analysis
is rather involved \cite{PrB}, we simply present the most important results 
as illustrated by the renormalization group flow lines in the 
$\sigma_{xx}$ and $\sigma_{xy}$ conductance plane in Fig. 2. 

\begin{figure}[t]
  \centerline{
  \epsfbox{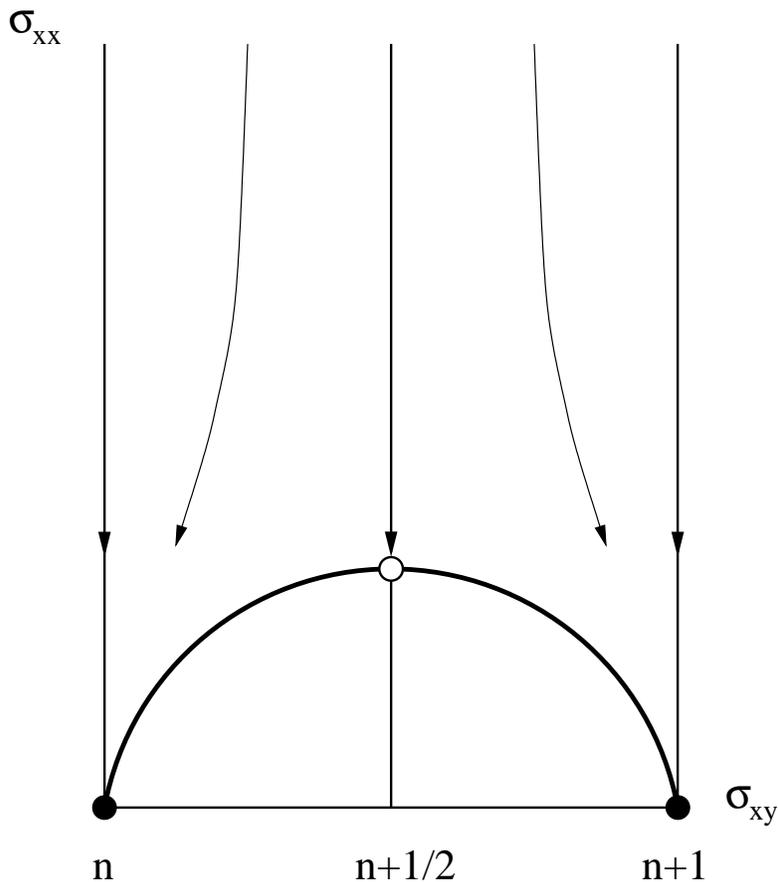}}
\caption{The renormalization group flow for the conductances. The arrows indicate the scaling towards 
the infrared.}
\end{figure}

The general consequences of the Coulomb interactions for the quantum critical behavior of the plateau
transitions will be reported elsewhere. On the other hand, there is the 
problem of {\em robust} quantization of the Hall conductance which
is represented in Fig. 2 by the strong coupling fixed points at integer values of
$\sigma_{xy}$. This aspect of the theory can obviously not be 
obtained from any analysis in the weak coupling regime, either perturbative or non-perturbative,
and it generally requires a more explicit knowledge of physics of incompressible the quantum Hall states.

Recently, a new and general ingredient of the instanton vacuum has been discovered
from which the phenomenon of robust quantization can be derived. It has turned out that the edge of
the instanton vacuum is generally {\em massless} and the theory can be mapped directly
onto the more familiar theory of chiral edge bosons in quantum Hall systems \cite{mishIII}.
The effective action for massless edge excitations, along with a mass gap for the excitations
in the bulk, is {\em dynamically} generated by the Finkelstein theory and quantized Hall
conductance now appears as the renormalized quantity $\sigma_{xy}^{\prime}$ in the 
effective action for the edge. 

It is important to remark that by extending the instanton vacuum approach to include
the effective action for edge excitations in the quantum Hall state \cite{mishIII}, the significance 
of $\cal F$ invariance has now also been demonstrated in the otherwise forbidden strong 
coupling regime of the Finkelstein theory.

\section{Chern Simons statistical gauge fields}

The theory of composite fermions \cite{qhecf} is obtained by adding the Chern Simons statistical gauge 
fields to 
the theory. The idea has been discussed at great length and in extensive detail at different 
places elsewhere. Here, we just mention how the flux attachement transformation by the Chern Simons
gauge fields generates a scaling diagram that includes both the abelian quantum Hall states and
the half integer effect (Fig. 3)\cite{mishI}. 
The theory also includes the Luttinger liquid theory for
edge excitations \cite{fracedge}. This, then, leads to a unified theory of both the compressible and 
incompressible states in the quantum Hall regime. 
\begin{figure}[h]
  \centerline{
   \epsfbox{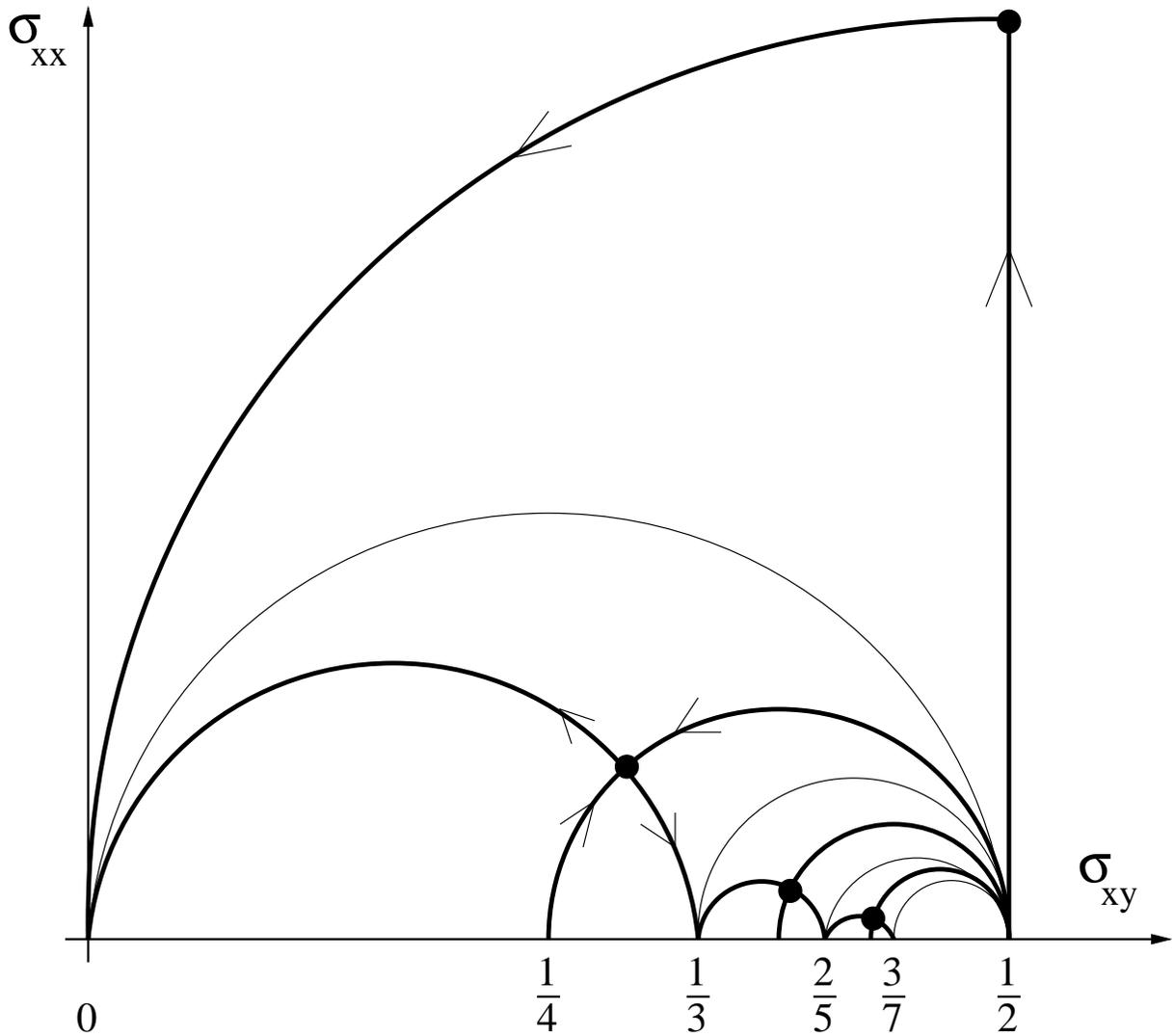}}
\caption{Unifying renormalization group diagram for integral and fractional quantum Hall effects.}
\end{figure}

\section{Acknowledgements}

This research was supported in part by INTAS (Grant No. 99-1070) and by the
Dutch Science Foundation FOM.


\begin{thebibliography}{99}

\bibitem{mishI} A.M.M. Pruisken, M.A. Baranov and B. \v{S}kori\'{c},
{\it Phys. Rev.} {\bf B60}, 16807 (1999). 

\bibitem{mishII} M.A. Baranov, A.M.M. Pruisken and B. \v{S}kori\'{c},
{\it Phys. Rev.} {\bf B60}, 16821 (1999). 

\bibitem{mishIII} A.M.M. Pruisken, B. \v{S}kori\'{c} and M. Baranov, 
{\it Phys. Rev.} {\bf B60}, 16838 (1999). The discussion in this paper
on the experiments on current scaling is somewhat controversial. 
See also \cite{PrB2}.

\bibitem{PrB} A.M.M. Pruisken and M.A. Baranov, {\it Europhys. Lett.}
{\bf 31} 543 (1995).

\bibitem{fp} A.M.M. Pruisken, {\it Nucl. Phys. B} {\bf 235} 277
(1984); H. Levine, S. Libby and A.M.M. Pruisken,
Phys. Rev. Lett. {\bf 51}, 20 (1983);
Nucl. Phys. {\bf B240} [FS12] 30, 49, 71 (1984). 

\bibitem{Fink} A.M. Finkelstein, Pis'ma Zh. Eksp. Teor. Fiz. {\bf 37}, 
436 (1983)
[JETP Lett. {\bf 37}, 517 (1983)]; 
Zh. Eksp. Teor. Fiz. {\bf 86}, 367 (1984) [Sov. Phys. JETP {\bf 59}, 
212 (1984)];
Physica B {\bf 197}, 636 (1994).

\bibitem{exp} A.M.M. Pruisken, Phys. Rev. Lett. {\bf 61}, 1297 (1988);
H.P. Wei, D.C. Tsui, M.A. Palaanen and
A.M.M. Pruisken, Phys. Rev. Lett. {\bf 61}, 1294 (1988); for recent results
see R.T.F. van Schaijk et al, Phys. Rev. Lett. {\bf 84},1567 (2000).

\bibitem{qhecf} S. Das Sarma ans A. Pinczuk, eds., {\it Perspectives in 
Quantum Hall Effects}
(John Wiley \& Sons,Inc.) 1997.
O. Heinonen, ed., {\it Composite Fermions} (World Scientific) 1998.

\bibitem{fracedge} B. \v{S}kori\'{c} and A.M.M. Pruisken, 
{\it Nucl. Phys. B} {\bf 559}, 637 (1999) and references therein.

\bibitem{PrB2} A.M.M. Pruisken and M.A. Baranov in preparation.

\bibitem{inst} A.M.M. Pruisken, Phys. Rev. {\bf B31} 416 (1985);
 A.M.M. Pruisken, Nucl. Phys. B {\bf 285} 719 (1987), {\bf 290} 61 (1987).


\end{thebibliography}
\end{document}